\documentclass[twocolumn,preprintnumbers,amsmath,amssymb,pra]{revtex4}
\usepackage{amsmath,amssymb,graphicx,epstopdf,bm,soul,upgreek, hyperref,natbib}
\usepackage{graphicx}
\usepackage{color}
\definecolor{OliveGreen}{rgb}{0,0.6,0}
\usepackage{dcolumn}
\usepackage{bm}
\usepackage{amssymb,float}
\usepackage{cancel}
\usepackage{ulem}
\usepackage{siunitx}

\begin{document}
\title{Non-reciprocal transport of Exciton-Polaritons in a non-Hermitian chain}

\author{S. Mandal$^1$}\email[Corresponding author:~]{subhaska001@e.ntu.edu.sg}
\author{R. Banerjee$^1$}
\author{Elena A. Ostrovskaya$^2$}
\author{T.C.H. Liew$^1$}\email[Corresponding author:~]{tchliew@gmail.com}

\affiliation{$^1$Division of Physics and Applied Physics, School of Physical and Mathematical Sciences, Nanyang Technological University, Singapore 637371, Singapore\\
$^2$ARC Centre of Excellence in Future Low-Energy Electronics Technologies and Nonlinear Physics Centre, Research School of Physics, The Australian National
University, Canberra, ACT 2601, Australia}

\begin{abstract}
We consider exciton-polaritons in a zigzag chain of coupled elliptical micropillars subjected to incoherent excitation. The driven-dissipative nature of the system along with the naturally present polarization splitting inside the pillars gives rise to  non-reciprocal dynamics,  which eventually leads to the non-Hermitian skin effect, where all the modes of the system collapse to one edge. As a result, the polaritons  propagate only in one direction along the chain, independent of the excitation position, and the propagation in the opposite direction is suppressed. The system shows robustness against  disorder and, using the bistable nature of  polaritons to encode information, we show one-way information transfer. This paves the way for compact and robust feedback-free one dimensional polariton transmission channels without the need for external magnetic field, which are compatible with proposals for polaritonic circuits.
\end{abstract}

\maketitle

{\textit{Introduction.}---} 
Non-reciprocal elements, where the transfer of a signal is favoured only in one direction \cite{Caloz_2018}, are an essential part of  information processing \cite{Keyes_1985}. However, designing such components in optical circuits is far from trivial due to the time-reversal invariance of  Maxwell's equations.  Magnetic materials  can be used to achieve on-chip optical isolation but they require large external magnetic fields \cite{Bi_2011}. Other ways to achieve optical non-reciprocity, such as time-varying fields and optical nonlinearity are  difficult to scale down to the microscale.

Optical information processing has been a particularly prominent topic in the field of exciton-polaritons, where one aims to benefit from the hybridization of an otherwise photonic system with a significant electronic nonlinearity.  Polaritonic switches \cite{Grosso_2014,Dreismann_2016,Lewandowski_2017}, transistors \cite{Lewandowski_2017,Gao_2012,Ballarini_2013,Zasedatelevi_2019}, amplifiers \cite{Wertz_2012,Niemietz_2016}, memories \cite{Ma_2017,Ma_2020}, and routers \cite{Flayac_2013, Marsault_2015, Schmutzler_2015} have already been realized. However, the future of this field depends on the development of mechanisms to connect elements without feedback.  Here the issue of how to generate non-reciprocal behaviour is as non-trivial as  in other optical systems.
\begin{figure}[t]
\includegraphics[width=0.5\textwidth]{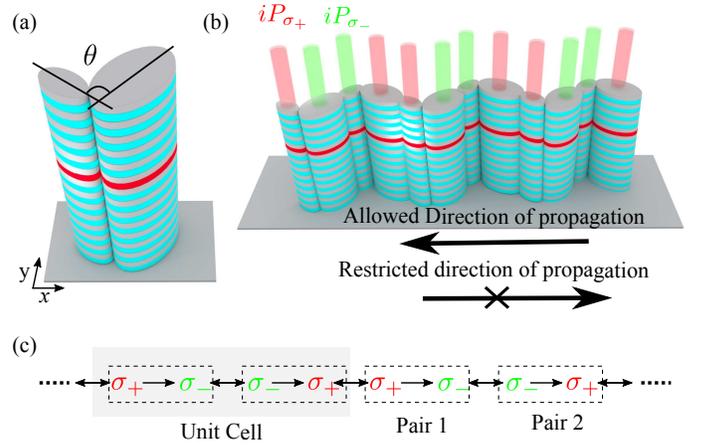}
\caption{Scheme: (a) A pair of coupled elliptical micropillars of different sizes; their mutual orientation introduces the polarization splitting angle $\theta$. (b) Micropillar pairs are arranged in a 1D zig-zag chain, where each micropillar is subjected to an incoherent excitation. (c) Effective coupling mechanism between the micropillars in the lattice. Each pair of micropillars indicated by the dashed boxes has nonreciprocal coupling. The unit cell of the 1D lattice is shown by the gray box.}
\label{Fig1}
\end{figure}

Significant motivation has been drawn from  topological photonics, where chiral edge states have been suggested for directionally dependent connections \cite{Lu_2014,Ozawa_2019}. Theoretically, robust polaritonic edge states can be achieved in many ways: by lifting the spin degeneracy and using spin-orbit coupling \cite{Karzig_2015,Bardyn_2015,Nalitov_2015,Gulevich_2017,Kartashov_2017,Li_2018,Sun_2019,Sigurdsson_2019}; by realizing Hofstadter's butterfly \cite{Banerjee_2018}; by using nonlinear interactions \cite{Bardyn_2016,Sigurdsson_2017}; by using staggered honeycomb lattices  \cite{Bleu_2018}; or by Floquet engineering \cite{Ge_2017}. The scheme in Refs.~\cite{Bardyn_2015,Nalitov_2015} has been realized experimentally under high external magnetic field using superconducting coils \cite{Klembt_2018}. Apart from the  bulky nature of this system, chiral edge states in this scheme appear in counter-propagating pairs, which can cause unwanted feedback. To circumvent this problem, mechanisms to switch off one of the edge states \cite{Mandal_Banerjee_2019}, to make both the edge states co-propagating  \cite{Mandal_2019}, or to use the polariton lifetime for input/output isolation \cite{Solnyshkov_2018}, have been considered. However, topological schemes remain ultimately inefficient for information transport. While plasmonic or exciton-polaritonic zigzag Su-Schrieffer-Heeger (SSH) chains \cite{Poddubny_2014,Jean_2017} demonstrate robust topological protection of edge-localised states, they show no transport. In 2D systems, chiral topological states can propagate along the edge of the system, however, they require a large-sized bulk to separate their edges for any chance of non-reciprocity.

Here, we propose a scheme for non-reciprocal propagation of exciton-polaritons in a quasi-1D geometry, which relies on inherent non-Hermiticity of the system. We make use of the recent experimental development of elliptical micropillars \cite{Klaas_2019,Gerhardt_2019}, where polarization splitting can be controllably engineered, and use the driven-dissipative nature of the system to demonstrate  propagation in a chain of micropillars even in the presence of disorder. We derive an effective $2\times2$ matrix Hamiltonian describing the coupling between two different polarized modes in neighbouring micropillars, which is mediated via other polarized modes. In a Hermitian system the coupling would necessarily be reciprocal. However, our effective Hamiltonian is non-Hermitian. Consequently, one of the off-diagonal elements of the matrix can vanish while the other doesn't, which is equivalent to a non-reciprocal coupling between two modes. In a chain of pillars, this eventually leads to the non-Hermitian skin effect, where the usual bulk-boundary condition breaks down and all the modes of the system become localised at one edge of the chain \cite{PRB.97.121401R.2018,PRL.124.056802.2020,JPhysMater.3.014002.2019,H_Lee_2019,Yao_2018,Ghatak_2019,Xiao_2020}. One of the key advantages of the system is that, robust propagation of polaritons  through a one dimensional chain of micropillars is obtained without the need for an external magnetic field. This  compact mechanism of non-reciprocal coupling is particularly promising  for future polaritonic devices. We also demonstrate the transport of binary information between localized sites where nonlinearity enables a bistable behaviour.

{\textit{The model.}---} We start by considering exciton-polariton states in a pair of coupled elliptical micropillars (Fig.~\ref{Fig1}(a))  described by the following set of driven dissipative Schr$\ddot{o}$dinger equations in the tight-binding limit:
\begin{align}
&i\hbar\frac{\partial\psi^l_{\sigma_+}}{\partial t}= \left(\varepsilon + {d\varepsilon}/{2}\right) \psi^l_{\sigma_+} +J\psi^r_{\sigma_+} +\Delta_T e^{+ 2i\theta_l} \psi^l_{\sigma_-},\label{Eq2}\\
&i\hbar\frac{\partial\psi^l_{\sigma_-}}{\partial t}=\left(\varepsilon + {d\varepsilon}/{2}-i\Gamma\right)\psi^l_{\sigma_-}+ J\psi^r_{\sigma_-} +\Delta_T e^{- 2i\theta_l} \psi^l_{\sigma_+},\label{Eq3}\\
&i\hbar\frac{\partial\psi^r_{\sigma_+}}{\partial t}=\left(\varepsilon -{d\varepsilon}/{2}-i\Gamma\right)\psi^r_{\sigma_+}+ J\psi^l_{\sigma_+} +\Delta_T e^{+ 2i\theta_r} \psi^r_{\sigma_-},\label{Eq4}\\
&i\hbar\frac{\partial\psi^r_{\sigma_-}}{\partial t}=\left(\varepsilon - {d\varepsilon}/{2}\right)\psi^r_{\sigma_-}+ J\psi^l_{\sigma_-} +\Delta_T e^{- 2i\theta_r} \psi^r_{\sigma_+}. \label{Eq5}
\end{align}
Here exciton-polariton modes in  each pillar are described by a wave function having  two circular polarization components, $\psi_{\sigma_\pm}$, where $l$ and $r$ represent the left and right pillars, respectively. We allow the modes in the left and right pillars to have different energies, where $\varepsilon$ is their average energy and $d\varepsilon$ is their energy difference. $\Gamma$ is the dissipation due to the finite lifetime of polaritons, which is compensated  \cite{Wertz_2012} using an incoherent excitation for modes $\psi^l_{\sigma_+}$ and $\psi^r_{\sigma_-}$. Each pillar is coupled to its neighbour by the Josephson  coupling term $J$. $\Delta_T$ is the polarization splitting inside each pillar which is naturally present in elliptical micropillars \cite{Gerhardt_2019} and $\theta$ represents the angle of polarization splitting equivalent to the micropillar orientation \cite{Supp}. It is helpful to  shift to a rotating frame by redefining the wavefunctions $\psi \rightarrow \psi \exp(-i \varepsilon t/\hbar)$ such that the effective onsite energies become $\pm d\varepsilon/2$. Due to the presence of $\Gamma$, the dynamics of $\psi^l_{\sigma_-}$ and $\psi^r_{\sigma_+}$ is  much faster compared to that of $\psi^l_{\sigma_+}$ and $\psi^r_{\sigma_-}$. As a result, on the timescale of $\psi^l_{\sigma_+}$ and $\psi^r_{\sigma_-}$, $\psi^l_{\sigma_-}$ and $\psi^r_{\sigma_+}$ can be approximated as stationary states
 \begin{align}
 &\psi^l_{\sigma_-}=-\frac{J\psi^r_{\sigma_-} +\Delta_T e^{- 2i\theta_l} \psi^l_{\sigma_+}}{(d\varepsilon/2-i\Gamma)},\label{Eq6}\\ 
&\psi^r_{\sigma_+}=-\frac{J\psi^l_{\sigma_+} +\Delta_T e^{+ 2i\theta_r} \psi^r_{\sigma_-}}{(-d\varepsilon/2-i\Gamma)}. \label{Eq7}
 \end{align}
Substituting  into Eqs.~(\ref{Eq2}) and (\ref{Eq5}), the  dynamics of the {\it slow} components of the system is given  by
\begin{equation}
i\hbar\frac{\partial}{\partial t}
\begin{pmatrix}
\psi^l_{\sigma_+} \\
\psi^r_{\sigma_-}
\end{pmatrix}
=\begin{pmatrix}
H_{ll}~ H_{lr} \\
H_{rl} ~H_{rr}
\end{pmatrix}
\begin{pmatrix}
\psi^l_{\sigma_+} \\
\psi^r_{\sigma_-}
\end{pmatrix},\label{Eq8}
\end{equation} 
where the effective Hamiltonian has the elements:
\begin{align}
&H_{ll}=  d\varepsilon/2-\frac{J^2}{(-d\varepsilon/2-i\Gamma)}-\frac{\Delta_T^2}{(d\varepsilon/2-i\Gamma)},\label{Eq9}\\
&H_{rr}= -d\varepsilon/2-\frac{J^2}{(d\varepsilon/2-i\Gamma)}-\frac{\Delta_T^2}{(-d\varepsilon/2-i\Gamma)},\label{Eq10}\\
&H_{lr}=-J\Delta_T\left[ \frac{e^{2i\theta_l}}{(d\varepsilon/2-i\Gamma)}+\frac{e^{2i\theta_r}}{(-d\varepsilon/2-i\Gamma)}\right],\label{Eq11}\\
&H_{rl}=-J\Delta_T\left[ \frac{e^{-2i\theta_l}}{(d\varepsilon/2-i\Gamma)}+\frac{e^{-2i\theta_r}}{(-d\varepsilon/2-i\Gamma)}\right].\label{Eq12}
\end{align}
Since this Hamiltonian is non-Hermitian ($H_{lr}\ne H_{rl}^*$) we can set  $H_{rl}=0$ while keeping $H_{lr}\ne0$. Indeed, such a condition can be obtained by setting \cite{Supp}
\begin{align}
d\varepsilon\neq 0,~~~\text{and} ~~~(\theta_r-\theta_l)=\arctan{\left(-{2\Gamma}/{d\varepsilon}\right)}. \label{NR1}
\end{align}
Note that according to Eqs. (\ref{Eq11}) and (\ref{Eq12}) it is essential that $\theta_l\neq\theta_r$,  which corresponds to neighbouring micropillars having different orientations, so that $H_{lr}\neq H_{rl}e^{4i\theta_l}$. The condition (\ref{NR1}) leads to non-reciprocal coupling between the micropillars. For the above mentioned excitation scheme,  $\sigma_+$ polaritons  hop from the left micropillar to the right one. It is also possible to do the same for the $\sigma_-$ polaritons by interchanging the components subject to incoherent excitation \cite{Supp}. 
\begin{figure}[t]
\includegraphics[width=0.45\textwidth]{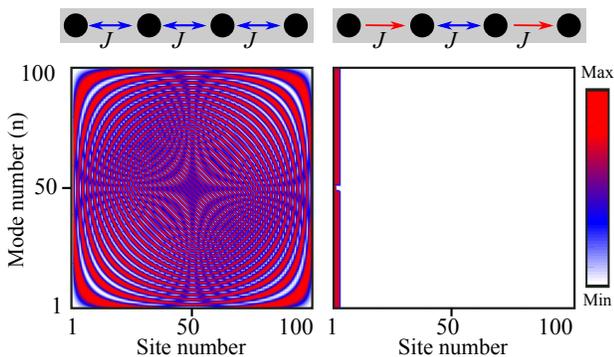}
\caption{Spatial profiles of different modes as a function of site number for a trivial chain of 100 micropillars (left panel) and non-reciprocal chain of 100 micropillars (right panel) in the tight binding limit. Here the blue arrows correspond to bidirectional coupling and the red arrows correspond to non-reciprocal coupling. Due to the non-Hermitian skin effect all the modes for the non-reciprocal chain collapse to the left edge. The most localized modes for $n=50$ and $51$ correspond to the zero energy modes, similar to those in Ref.~\cite{Comaron_arxiv}, but located at one edge. Parameter: $J=0.5$ meV.}
\label{Fig2_TB}
\end{figure}

The pairs of micropillars with non-reciprocal coupling can be combined into a chain (see Fig.~\ref{Fig1}(b)). Since the non-reciprocal transmission is also flipping the spin polarization, we use one pair to transport $\sigma_+$ polarization rightward to a $\sigma_-$ state, which is then coupled bidirectionally to another pair that transports the $\sigma_-$ state to a $\sigma_+$ state. This gives a unit cell of four micropillars (see Fig.~\ref{Fig1}(c)). Even though the connection between the pairs is bidirectional, the non-reciprocity within each pair ensures  non-reciprocity of the whole chain. In Fig.~\ref{Fig2_TB} the mode profiles of a trivial chain and the non-reciprocal chain are shown. For the trivial chain, the modes are distributed over all sites while for the case of the non-reciprocal chain  all the modes are located at one edge. This is known as the non-Hermitian skin effect \cite{Okuma_2020,Helbig_2020,Weidemann_2020}, which occurs in non-Hermitian systems with non-reciprocity  and can not be reproduced in Hermitian systems. To obtain the non-Hermitian skin effect, it is not necessary to fully switch off one of the non-diagonal terms; a small anisotropy between them is sufficient for collapsing the modes to one edge (see supplementary Movie 1).

{\textit{Band structure and pulse propagation.}---}  To illustrate that our results do not depend on the tight-binding approximation, we now move to a continuous model in space  using the driven-dissipative Gross-Pitaevskii equation, 
\begin{align}
i\hbar\frac{\partial\psi_{\sigma_\pm}}{\partial t}=&\Big[-\frac{\hbar^2\nabla^2}{2m}+V(x,y)+i\hbar\left(p_{\sigma_\pm}(x,y)-\gamma(x,y)\right)\nonumber\\
&+\hbar gp_{\sigma_\pm}(x,y)+I^{NL}_{\sigma_\pm}\Big]\psi_{\sigma_\pm}+V_T(x,y,\pm\theta)\psi_{\sigma_\mp}.\label{GPE}
\end{align}
Here $m$ is the effective polariton mass, $V(x,y)$ is the effective potential representing a zigzag chain of elliptical micropillars (see Fig.\ref{Fig1}(b)), $p_{\sigma_\pm}$ and $\gamma$ characterise the rate of injection of polaritons by the incoherent pump and the rate of polariton decay, respectively. $\gamma$ outside the micropillars is two times larger than the inside. The term with the dimensionless  $g$ factor is introduced to take into account the potential created by the excitonic reservoir, which has the same profile as the incoherent excitation. $I^{NL}_{\sigma_\pm}=\left(\alpha_1-i\alpha_{NL}\right)|\psi_{\sigma_\pm}|^2+\alpha_2 |\psi_{\sigma_\mp}|^2$ is the nonlinear contribution, which includes: interactions between polaritons with the same spin (with strength $\alpha_1$), opposite spin (with strength $\alpha_2$), and gain saturation caused by the depletion of the excitonic reservoir (with strength $\alpha_{NL}$) \cite{Borgh_Keeling}. $2V_T$ represents the polarization splitting inside the micropillars,  which is modelled with the same spatial profile as $V$ with an extra factor of $\exp(\pm 2i\theta)$  to take into account the orientation of each pillar \cite{Supp}. In writing Eq.~(\ref{GPE}), we have assumed that the dynamics of the reservoir is fast and therefore  its effect can be modelled by the effective gain, effective potential, and  the gain saturation ($\alpha_{NL}$) terms.  Our results are independent  of this assumption \cite{Supp} and can also be reproduced with an explicit account of reservoir dynamics \cite{Wouters_Carusotto_2007}.

\begin{figure}[t]
\includegraphics[width=0.45\textwidth]{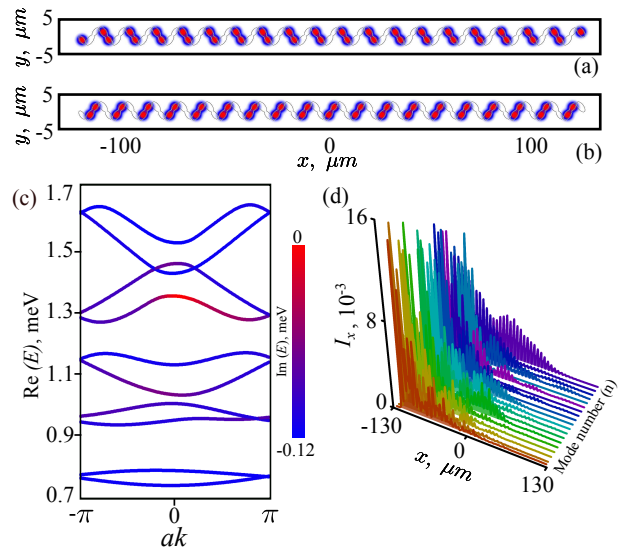}
\caption{(a-b) The spatial profile of the incoherent pumps $P_{\sigma_+}$ and  $P_{\sigma_-}$, respectively,  composed of  Gaussians of width $2.1$ $\mu$m positioned at the center of the relevant micropillars. Such an arrangement is within the reach of experimental technology \cite{Topfer_2020}. (c)  Bandstructure of an infinite chain under incoherent excitation. The states shown in red correspond to the {\it slow} modes and decay slower than all other states shown in blue.  (d) Spatial profile, $I_x=\sum_{\sigma_\pm}\int|\psi_\sigma(x,y)|^2~dy$, corresponding to the {\it slow} modes for the case of a finite chain. Here different colours denote different modes.}
\label{Fig2}
\end{figure}
The incoherent pumps are arranged so that the  {\it slow} component of the polariton mode in each micropillar has almost zero decay.  The spatial profile of the incoherent pump is shown in Figs.~\ref{Fig2}(a-b). Now, we have all the ingredients to calculate the band structure of the linear system ($I^{NL}_{\sigma_\pm}=0$) under the periodic boundary condition, which is shown in Fig.~\ref{Fig2}(c). Since there are four sites in one unit cell, the real part of the low energy band structure of the system is composed of four bands. It should be noted that this is an exact band structure of the system without the approximations used in Eqs.~(\ref{Eq6}-\ref{Eq7}) and the states having lower decay correspond to the {\it slow} modes in the approximated Hamiltonian in Eq.~(\ref{Eq8}). Each state in the real part of the band structure is color coded according to the imaginary part with red corresponding to the lowest decay ({\it slow} modes), and blue corresponding to the larger decay. The nonreciprocal nature of the system can be attributed to the fact that,  the states shown in red (see the lower branch of the fourth band in Fig.~\ref{Fig2}(c)) will accumulate the polaritons relaxing from higher energy. These states have a negative group velocity, $v_g=({\partial E}/{\partial k})/\hbar$; whereas more lossy states have positive group velocity. Overall, we can expect to have polariton propagation at a speed  $v_g$ along only one direction in the micropillar chain. The spatial profiles of highly occupied modes plotted in Fig.~\ref{Fig2}(d) are in agreement with the modes obtained in the tight binding limit shown in Fig.~\ref{Fig2_TB}. Although the bulk is translationally symmetric, the localization of the modes  at the left edge of the chain indicates the break down of the usual bulk-boundary correspondence in Hermitian systems \cite{Helbig_2020}. 
\begin{figure}[t]
\includegraphics[width=0.48\textwidth]{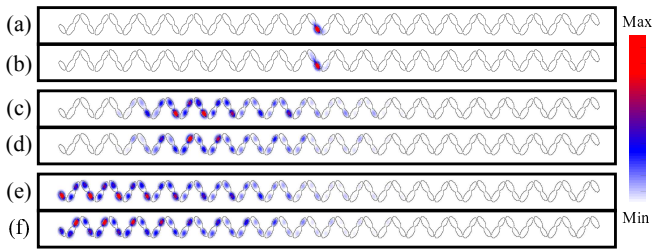}
\caption{(a, c, e) Density  of the $\sigma_+$ polaritons under an incoherently pulsed excitation for 10 ps, 90 ps, and 180 ps, respectively. The same for the $\sigma_-$ polaritons are shown in (b, d, f). The non reciprocal nature of the system is clear from the fact that the polaritons propagate only in one direction while the propagation in the opposite direction is suppressed. The width of the incoherent pulse is taken as 2.1 $\mu$m and the amplitude is taken as 5$p_0$, where $p_0=0.68$ ps$^{-1}$ \cite{Parameters_Used}.}
\label{Fig3}
\end{figure}

The above band structure calculation is performed by using  typical experimental parameters \cite{Parameters_Used}. For the angle between the two micropillars inside one micropillar pair around \ang{110}, the periodicity of the lattice along the $x$ direction becomes $a=12.9~\mu$m, which corresponds to $v_g=-0.7~\mu $m/ps. The negative value of $v_g$ means that the polaritons propagate from right to left along the micropillar chain.

To demonstrate the non-reciprocal polariton propagation, we apply a Gaussian shaped incoherent excitation pulse in the middle of the chain. The dynamics of the polaritons can be seen in Fig.~\ref{Fig3}. Unlike a trivial chain of micropillars, where the polaritons propagate  in both directions from the excitation spot \cite{Supp}, in this case they propagate only in one direction. As explained above, this is due to the  states with lower losses acquiring a larger polariton population compared to the decaying states.  Remarkably, all the polaritons in the system will be localised at the left edge of the chain, regardless of the position of the excitation spot. This is quite similar to the recently realized topological funneling of light \cite{Weidemann_2020}. From the intensity profile it is clear that only the {\it slow} components of the on-site modes get excited, and they are mostly localized in the micropillars with smaller dimension, which can be attributed to  the dominant contribution of the smaller pillars to  fourth band in the bandstructre (see Fig.~\ref{Fig2}(c))  (see the Bloch states in \cite{Supp}).

\begin{figure}[t]
\includegraphics[width=0.45\textwidth]{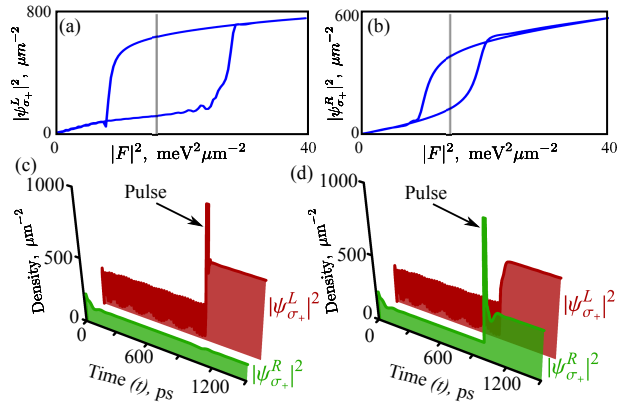}
\caption{(a-b) Bistability curves of the pillars at the left and right end of the chain, respectively. The energy of the resonant pump, $F$,  is fixed at 1.38 meV. The gray line in the bistability curves represents the pump values ($F=4.1$ meV$\mu$m$^{-1}$ and $3.9$ meV$\mu$m$^{-1}$, respectively, for left and right end pillars) used to demonstrate the switching. (c-d) Time dynamics of the end pillars when the incoherent pulse is positioned at the left and right end, respectively. Pulse positioned at the right end induces switching at both ends, whereas the pulse positioned at the left end does not switch the pillar at the right end. The spatial profile of the resonant pump and incoherent pulse are the same as the one in Fig~\ref{Fig3}. The peak value of the pulse is taken as 35$p_0$. To visualize the switching effect the maximum density is cut at  $10^3~\mu$m$^{-2}$.}
\label{Fig4}
\end{figure}
{\textit{Demonstration of feedback suppression.}---} Due to their strong nonlinearity polaritons show bistable behavior which makes them suitable for several  applications, such as solving NP-hard problems  \cite{Kyriienko_2019}, realizing bistable topological insulators \cite{Kartashov_Bis_2017,Banerjee_2020},    universal logic gates \cite{Ortega_2013} and enhancement of dark soliton stability \cite{Lerario_2003}. In this section we introduce additional resonant pumps placed at the two ends of the chain such that the end pillars, with the on-site wave-functions $\psi^L_{\sigma_+}$ and $\psi^R_{\sigma_+}$, are initially in their  lower bistable states. The bistability curves of the end pillars, plotted in Fig.~\ref{Fig4} (a-b), are obtained by slowly varying the pump, $F$. Next, an incoherent pulse is introduced at the left end of the chain, which switches $\psi^L_{\sigma_+}$ from its lower bistable state to the upper one. Due to the non-reciprocal nature of the system, polaritons can not propagate from left to right and therefore no switching is observed for $\psi^R_{\sigma_+}$ (see Fig.~\ref{Fig4} (c)). Then, we introduce the same incoherent pulse at the right end of the chain. As expected, $\psi^R_{\sigma_+}$ switches instantly, but more importantly $\psi^L_{\sigma_+}$ also switches after some time (see Fig.~\ref{Fig4} (d)). This can be thought of as feedback suppressed information processing, where the information is encoded in the bistability and transmitted in one direction only (from right to left). The nonlinear coefficients used for the calculations are $\alpha_1=1~\mu$eV$\mu$m$^2$ \cite{Interaction_Const}, $\alpha_2=-0.05\alpha_1$ \cite{Kavokin_Renucci_2005,Vladimirova_Cronenberger_2010,Lecomte_Taj_2014}, and $\alpha_{NL}=0.3\alpha_1$ \cite{Keeling_2008}. Although we have modelled the scheme with resonant excitation, we expect that the non-reciprocal transport mechanism would also be compatible with  bistability under non-resonant excitation\cite{Karpov_2015,Ohadi_Dreismann_2015,Zhang_2017}. 

{\textit{Discussion and conclusion.}---}  We have presented a scheme for  non-reciprocal exciton-polariton transport in a quasi-1D chain of elliptical micropillars without an external magnetic field.  Due to the non-reciprocal coupling within  micropillar pairs, all the highly populated polariton states  are localized at one edge of the  chain. This makes the polaritons  propagate in one direction  along the chain regardless of the excitation position. This non-reciprocity also protects against backscattering (see \cite{Supp} for the discussion on the robustness against disorder) and allows one-way information transfer. While our theory is restricted to the semi-classical regime, it would be interesting to extend it to the quantum optical regime, where non-reciprocal blockade effects are anticipated \cite{Shen_2020}. Due to its compactness, such a chain can be extremely useful in connecting different components of future polaritonic circuits such as the polariton neural networks \cite{Opala_2019,Ballarini_2020}.              

{\textit{Acknowledgment.}---} The work was supported by the Ministry of Education, Singapore (Grant No. MOE2019-T2-1-004) and the Australian Research Council (ARC). T. L. thanks E. Z. Tan for discussions.

\end{document}


\title{Supplemental material for \\

Non-reciprocal transport of Exciton-Polaritons in a non-Hermitian chain}%
\author{S. Mandal$^1$}\email[Corresponding author:~]{subhaska001@e.ntu.edu.sg}
\author{R. Banerjee$^1$}
\author{Elena A. Ostrovskaya$^2$}
\author{T.C.H. Liew$^1$}\email[Corresponding author:~]{tchliew@gmail.com}

\affiliation{$^1$Division of Physics and Applied Physics, School of Physical and Mathematical Sciences, Nanyang Technological University, Singapore 637371, Singapore\\
$^2$ARC Centre of Excellence in Future Low-Energy Electronics Technologies and Nonlinear Physics Centre, Research School of Physics, The Australian National
University, Canberra, ACT 2601, Australia}


\maketitle
\begin{figure}[H]
\centering
\includegraphics[width=0.25\textwidth]{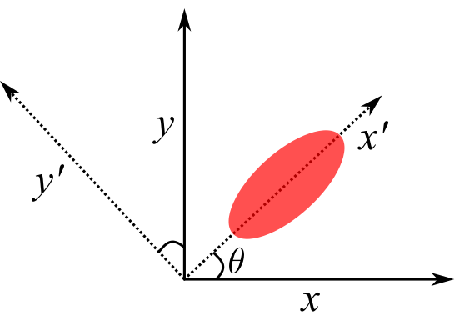}
\caption{A schematic diagram of an elliptical micropillar rotated by an angle $\theta$.}
\label{Supp1}
\end{figure}

{\textit{Realization of the splitting angle.}---} Let us consider an elliptical micropillar rotated by an angle $\theta$ as shown in Fig.~\ref{Supp1}. Modes having linear polarization along $x^{\prime}$ and $y^{\prime}$ will have a splitting in energy $\Delta_T$ due to the shape anisotropy. In the pump-loss free picture, the dynamics of the polaritons in the primed basis can be represented by 
\begin{equation}
i\hbar\frac{\partial}{\partial t}\begin{pmatrix}
\psi^\prime_{x} \\
\psi^\prime_{y} 
\end{pmatrix}
=\begin{pmatrix}
\varepsilon_{x}^{\prime}~ ~~~0\\
0 ~~~~\varepsilon_{y}^{\prime}
\end{pmatrix}
\begin{pmatrix}
\psi^\prime_{x} \\
\psi^\prime_{y} 
\end{pmatrix},
\end{equation} 
where, $\Delta_T=\varepsilon_{x}^{\prime}-\varepsilon_{y}^{\prime}$; and $\varepsilon_{x(y)}^{\prime}$  is the eigen-energy of the mode $\psi^\prime_{x(y)}$. In order to move to the $x$ and $y$ polarization basis we make the following transformation
\begin{equation}
\begin{pmatrix}
\psi^\prime_{x} \\
\psi^\prime_{y} 
\end{pmatrix}
=\begin{pmatrix}
~~\cos({\theta})~ ~~~\sin(\theta)\\
-\sin(\theta) ~~~~\cos({\theta})
\end{pmatrix}
\begin{pmatrix}
\psi_{x} \\
\psi_{y} 
\end{pmatrix}.
\end{equation} 
Modes in the circular polarization basis are related to those in the linear polarization basis by
\begin{equation}
\begin{pmatrix}
\psi_{\sigma_+} \\
\psi_{\sigma_-} 
\end{pmatrix}
=\frac{1}{\sqrt{2}}\begin{pmatrix}
1~ ~~~i\\
1 ~-i
\end{pmatrix}
\begin{pmatrix}
\psi_{x} \\
\psi_{y} 
\end{pmatrix},
\end{equation} 
or,
\begin{equation}
\begin{pmatrix}
\psi_{x} \\
\psi_{y} 
\end{pmatrix}
=\frac{1}{\sqrt{2}}\begin{pmatrix}
~~1~ ~~~1\\
-i~~~ ~i
\end{pmatrix}
\begin{pmatrix}
\psi_{\sigma_+} \\
\psi_{\sigma_-} 
\end{pmatrix}.
\end{equation} 
The final transformation matrix for going to the circular polarization basis from the primed basis becomes
\begin{equation}
T
=\frac{1}{\sqrt{2}}\begin{pmatrix}
~~\cos({\theta})~ ~~~\sin(\theta)\\
-\sin(\theta) ~~~~\cos({\theta})
\end{pmatrix}\begin{pmatrix}
~~1~ ~~~1\\
-i~~~ ~i
\end{pmatrix}=\frac{1}{\sqrt{2}}\begin{pmatrix}
~~e^{-i\theta}~ ~~~e^{i\theta}\\
-ie^{-i\theta}~~~ ~ie^{i\theta}
\end{pmatrix}
\end{equation} 
Finally, the Hamiltonian in the circular polarization basis can be represented by
\begin{equation}
H_{\sigma_{\pm}}=T^{-1}\begin{pmatrix}
\varepsilon_{x}^{\prime}~ ~~~0\\
0 ~~~~\varepsilon_{y}^{\prime}
\end{pmatrix}T=\frac{1}{2}\begin{pmatrix}
\varepsilon_{x}^{\prime}+\varepsilon_{y}^{\prime}~ ~~~~~~~~~e^{2i\theta}(\varepsilon_{x}^{\prime}-\varepsilon_{y}^{\prime})\\
e^{-2i\theta}(\varepsilon_{x}^{\prime}-\varepsilon_{y}^{\prime}) ~~~~~~~\varepsilon_{x}^{\prime}+\varepsilon_{y}^{\prime}
\end{pmatrix}=\begin{pmatrix}
\varepsilon_{x}^{\prime}+\varepsilon_{y}^{\prime}~ ~~~~~~~~~e^{2i\theta}\Delta_T \\
e^{-2i\theta}\Delta_T~~~~~~~\varepsilon_{x}^{\prime}+\varepsilon_{y}^{\prime}
\end{pmatrix}\label{SEq1}
\end{equation}
It can be clearly seen that the splitting angle acts as a coupling phase between the two circularly polarized modes. 

\begin{figure}[H]
\centering
\includegraphics[width=0.6\textwidth]{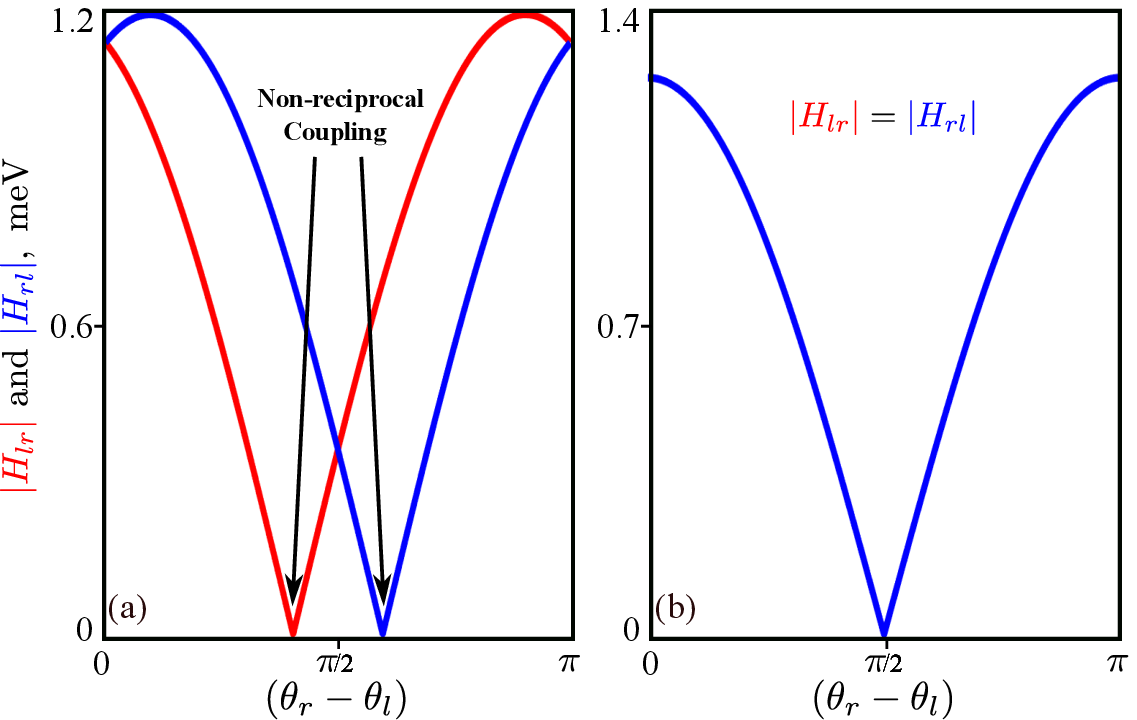}
\caption{Plot of $|H_{lr}|$ (in red) and $|H_{rl}|$ (in blue) as a function of $(\theta_r-\theta_l)$.  For $d\varepsilon\neq0$ the system shows anisotropic coupling (plotted in (a)), which becomes usual bidirectional coupling for $d\varepsilon=0$. Parameters: $J=0.5$ meV, $\Delta_T=0.2$ meV, $\Gamma=0.16$ meV, $d\varepsilon=0.1$ meV for (a) and $0$ meV for (b). }
\label{Supp1_2}
\end{figure}

{\textit{Derivation of the non-reciprocity condition Pair1.}---} Let us recall the off-diagonal elements of the effective coupling matrix given by Eqs. (10-11) of the main text:
\begin{align}
&H_{lr}=-J\Delta_T\left[ \frac{e^{2i\theta_l}}{(d\varepsilon/2-i\Gamma)}+\frac{e^{2i\theta_r}}{(-d\varepsilon/2-i\Gamma)}\right]\label{SEq2}\\
&H_{rl}=-J\Delta_T\left[ \frac{e^{-2i\theta_l}}{(d\varepsilon/2-i\Gamma)}+\frac{e^{-2i\theta_r}}{(-d\varepsilon/2-i\Gamma)}\right].
\end{align}
For $H_{rl}$ to vanish we require
 
\begin{align}
 \frac{e^{-2i\theta_l}}{(d\varepsilon/2-i\Gamma)}&=-\frac{e^{-2i\theta_r}}{(-d\varepsilon/2-i\Gamma)}\nonumber\\
\therefore e^{2i\left(\theta_r-\theta_l\right)}&=-\frac{(d\varepsilon/2-i\Gamma)}{(-d\varepsilon/2-i\Gamma)}=e^{2i\arctan{\left(-2\Gamma/d\varepsilon\right)}}\label{SEqR1_1}\\
\therefore \left(\theta_r-\theta_l\right)&=\arctan{\left(-2\Gamma/d\varepsilon\right)}\label{SEqR1_2}.
\end{align}
To illustrate the non-reciprocal coupling, we plot $|H_{lr}|$ and $|H_{rl}|$ as a function of the splitting angle in Fig.~\ref{Supp1_2}. The coupling is always anisotropic for $(\theta_r-\theta_l)\neq n\pi/2$, where $n$ can be zero or any integer. The non-reciprocal coupling conditions where one of the coupling terms goes to zero are indicated by the arrows in Fig.~\ref{Supp1_2}(a). The same quantities are plotted in Fig.~\ref{Supp1_2}(b) for  $d\varepsilon=0$, which shows the usual bidirectional coupling with $|H_{lr}|=|H_{rl}|$ for all values of $(\theta_r-\theta_l)$. This also shows the importance of $d\varepsilon$ in this scheme.

Substituting the condition for $H_{rl} = 0$ given by Eq.~(\ref{SEqR1_1}) into the equation for $H_{lr}$ in Eq.~(\ref{SEq2}), we find that: $|H_{lr}|^2=\left(16J\Delta_Td\varepsilon\Gamma\right)^2/\left(d\varepsilon^2+\Gamma^2\right)^3$. Although, technically for any finite value of $d\varepsilon$ one can obtain finite $H_{lr}$ when $H_{rl} = 0$, the size of $H_{lr}$ (which determines the strength of coupling in the desired direction) is dependent on the size of $d\epsilon$ and is optimal when $d\varepsilon=\sqrt{2}\Gamma$.

\newpage
{\textit{Derivation of the non-reciprocity condition for Pair2.}---} Here we exchange the components of the incoherent excitation and  the evolution of the modes now become
\begin{align}
&i\hbar\frac{\partial\psi^l_{\sigma_+}}{\partial t}= (\varepsilon+d\varepsilon/2-i\Gamma)\psi^l_{\sigma_+} +J\psi^r_{\sigma_+} +\Delta_T e^{+ 2i\theta_l} \psi^l_{\sigma_-},\\
&i\hbar\frac{\partial\psi^l_{\sigma_-}}{\partial t}=(\varepsilon+d\varepsilon/2)\psi^l_{\sigma_-}+ J\psi^r_{\sigma_-} +\Delta_T e^{- 2i\theta_l} \psi^l_{\sigma_+},\\
&i\hbar\frac{\partial\psi^r_{\sigma_+}}{\partial t}=(\varepsilon-d\varepsilon/2)\psi^r_{\sigma_+}+ J\psi^l_{\sigma_+} +\Delta_T e^{+ 2i\theta_r} \psi^r_{\sigma_-},\\
&i\hbar\frac{\partial\psi^r_{\sigma_-}}{\partial t}=(\varepsilon-d\varepsilon/2-i\Gamma)\psi^r_{\sigma_-}+ J\psi^l_{\sigma_-} +\Delta_T e^{- 2i\theta_r} \psi^r_{\sigma_+}. 
\end{align}
Next we move to the rotating frame  by redefining the wavefunctions, $\psi \rightarrow \psi \exp(-i \varepsilon t/\hbar)$, such that the onsite energies become $\pm d\varepsilon/2$. Here the dynamics of the modes $\psi^l_{\sigma_+}$ and $\psi^r_{\sigma_-}$ will be much faster (compared to those of $\psi^l_{\sigma_-}$ and $\psi^r_{\sigma_+}$) such that they can be approximated as steady states given by
 \begin{align}
 &\psi^l_{\sigma_+}=-\frac{J\psi^r_{\sigma_+} +\Delta_T e^{2i\theta_l} \psi^l_{\sigma_-}}{(d\varepsilon/2-i\Gamma)},\\ 
&\psi^r_{\sigma_-}=-\frac{J\psi^l_{\sigma_-} +\Delta_T e^{- 2i\theta_r} \psi^r_{\sigma_+}}{(-d\varepsilon/2-i\Gamma)}. 
 \end{align}
 
 In this case the non-diagonal terms of the effective Hamiltonian becomes
\begin{align}
&H_{lr}=-J\Delta_T\left[ \frac{e^{-2i\theta_l}}{(d\varepsilon/2-i\Gamma)}+\frac{e^{-2i\theta_r}}{(-d\varepsilon/2-i\Gamma)}\right],\\
&H_{rl}=-J\Delta_T\left[ \frac{e^{2i\theta_l}}{(d\varepsilon/2-i\Gamma)}+\frac{e^{2i\theta_r}}{(-d\varepsilon/2-i\Gamma)}\right].
\end{align}
Following the steps as those for Pair 1, the non reciprocity condition $H_{lr}\neq 0$ and $H_{rl}=0$; is obtained for $\left(\theta_r-\theta_l\right)=\arctan{\left(2\Gamma/d\varepsilon\right)}$. It should be noted that  to satisfy the condition on the angle, the orientation of the pillars in Pair 2 should be opposite to those in Pair 1.\\

{\textit{Supplementary movies.}---} In this section we have considered a chain of 100 elliptical micropillars using the aforementioned pairs as a repeating unit. All the parameters are kept the same as those in Fig.~\ref{Supp1_2}(a). The spatial profile of the modes as a function of $(\theta_r-\theta_l)$ corresponding to the slow Hamiltonian in the tight binding limit is shown in Movie 1.  Surprisingly, the modes of the system are always localized at a particular edge for all values of $(\theta_r-\theta_l)\neq n\pi/2$, where $n=0,1,2$.  Since $|H_{rl}|>|H_{lr}|$ for $0<(\theta_r-\theta_l)<\pi/2$, all the modes are localized at the right edge. For $\pi/2<(\theta_r-\theta_l)<\pi$,  the strength of the coupling becomes $|H_{lr}|>|H_{rl}|$, which shifts all the modes from the right edge to the left one. For $(\theta_r-\theta_l)=n\pi/2$, where $n=0,1,2$, there is no anisotropy between the coupling terms ($|H_{rl}|=|H_{lr}|$) and the modes are no longer localised at a particular edge. 

In Movie 2, to check the effect of the boundary condition, we calculate the spatial profile of the modes corresponding to the same chain as above but using the periodic boundary condition. For this case, all the modes have contribution from all the sites of the chain for all values of ($\theta_r-\theta_l$). 

In Movie 3, we once again calculate the spatial profile of the modes corresponding to the same chain as in Movie 1 but without the onsite term by putting $d\varepsilon=0$. As it can be seen from Fig.~\ref{Supp1_2}(b), this corresponds to usual bidirectional coupling and all the modes spread through all the sites of the chain. However, when $(\theta_r-\theta_l)$ is near $\pi/2$, $J>|H_{rl}|=|H_{lr}|$, and the system behaves as a Su-Schrieffer-Heeger (SSH) chain with modes localized at two edges. For $(\theta_r-\theta_l)=\pi/2$, $|H_{rl}|=|H_{lr}|=0$, which corresponds to isolated sites.

\begin{figure}[H]
\centering
\includegraphics[width=0.65\textwidth]{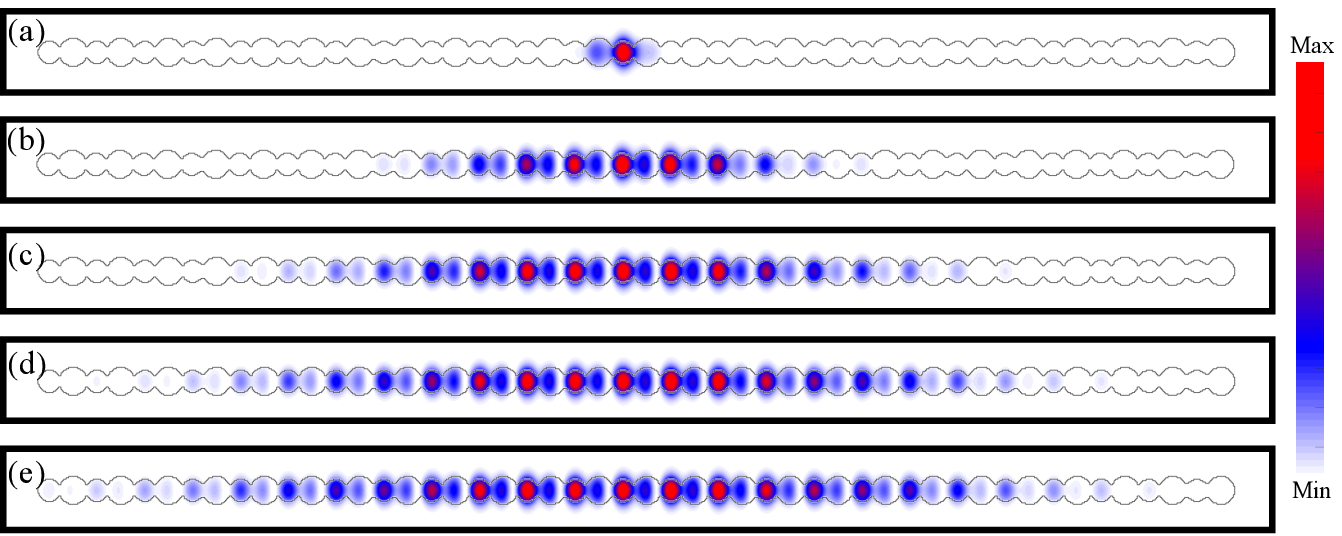}
\caption{Polariton propagation under an incoherent pulsed excitation through a straight chain of micropillars where the reciprocity is not broken. From the excitation spot the polaritons propagate equally in both directions. (a-e) Intensity of the polaritons for 10 ps, 50 ps, 100 ps, 150 ps and 200 ps, respectively.}
\label{Supp2}
\end{figure}

\newpage
{\textit{Pulse propagation in a system with reciprocity.}---} 
In this section we show  propagation of an  incoherent pulse in a system with reciprocity. Such a system can be easily prepared by considering a straight chain of micropillars instead of a zigzag one. The dynamics of the polaritons can be seen in Fig.~\ref{Supp2} where the total intensity of the polaritons is plotted. Unlike the non-reciprocal chain here polaritons propagate in both directions from the excitations spot.\\

{\textit{Spatial profiles of the slow decaying Bloch modes in Fig. 3 in the main text }---} In this section we plot the spatial profile of the slow decaying Bloch modes from the fourth band of the band structure shown in Fig. 3 in the main text. Similar to the periodic case in the tight binding model (see Movie 2) these modes are spatially distributed over  all sites of the chain. They also have their main intensity located at the smaller micropillars.
\begin{figure}[H]
\centering
\includegraphics[width=0.8\textwidth]{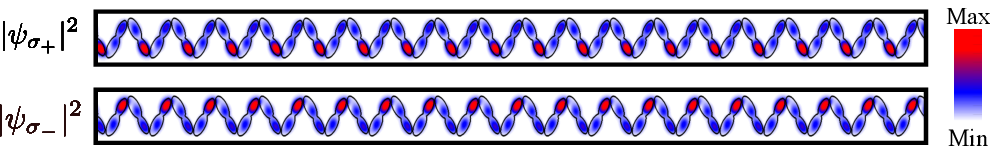}
\caption{Spatial profile of the slow decaying Bloch modes.}
\label{Supp2_1}
\end{figure}

\begin{figure}[H]
\centering
\includegraphics[width=0.98\textwidth]{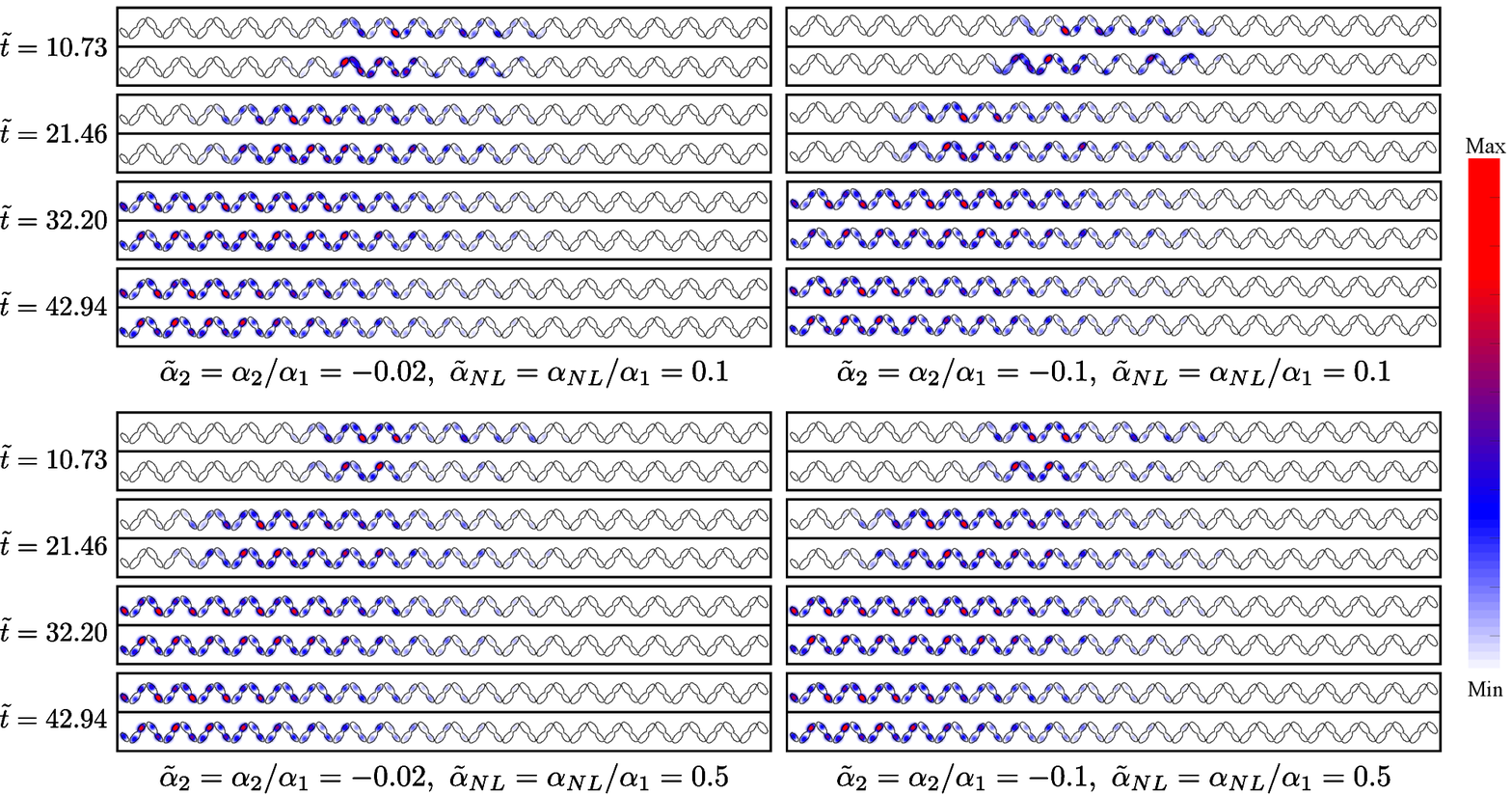}
\caption{Non-reciprocal propagation of polaritons considering different nonlinear coefficients. At each time step the top  and the bottom panel corresponds to $ |\tilde{\psi}_{\sigma_+}|^2$ and $ |\tilde{\psi}_{\sigma_-}|^2$, respectively. All other parameters correspond to those used in Fig.~4 in the main text.}
\label{SNonlinear_para_Dep}
\end{figure}

{\textit{Effect of the nonlinear coefficients on the non-reciprocal polariton propagation.}---} 
To check the influence of the nonlinear coefficients on the non-reciprocal propagation, we consider  both the cases where the nonlinear coefficients $\alpha_1, ~\alpha_2,$ and $\alpha_{NL}$ are larger and smaller than those considered in the main text. First we move to the dimensionless coordinate by introducing the following dimensionless variables: $\tilde{t}=t/T_u$, $ \tilde{p}=pT_u$, $ \tilde{\gamma}=\gamma T_u$, $\tilde{x}=x/L_u$, $ \tilde{y}=y/L_u$, $ \tilde{V}=V/E_u$ , $ \tilde{V}_T=V_T/E_u$, $\tilde{\psi}_{\sigma_{\pm}}=\sqrt{\alpha_1}\psi_{\sigma_{\pm}}/\sqrt{E_u}$, where $L_u=3~\mu$m is the length unit, $E_u=\hbar^2/2mL_u^2=0.14$ meV is the energy unit and $T_u=\hbar/E_u=4.65$ ps is the time unit. Using the dimensionless variables, Eq.~(13) of the main text can be expressed as
\begin{align}
i\frac{\partial\tilde{\psi}_{\sigma_\pm}}{\partial \tilde{t}}=&\Big[-\frac{\tilde{\nabla}^2}{2}+\tilde{V}(\tilde{x},\tilde{y})+i\left(\tilde{p}_{\sigma_\pm}(\tilde{x},\tilde{y})-\tilde{\gamma}(\tilde{x},\tilde{y})\right)+g\tilde{p}_{\sigma_\pm}(\tilde{x},\tilde{y})+\left(1-i\tilde{\alpha}_{NL}\right)|\tilde{\psi}_{\sigma_\pm}|^2+\tilde{\alpha}_{2}|\tilde{\psi}_{\sigma_\mp}|^2\Big]\tilde{\psi}_{\sigma_\pm}\nonumber\\
&+\tilde{V}_T(\tilde{x},\tilde{y},\pm\theta)\psi_{\sigma_\mp},\label{SGPE}
\end{align}
where $\tilde{\nabla}^2=-(\partial^2/\partial \tilde{x}^2+\partial^2/\partial \tilde{y}^2)$, $\tilde{\alpha}_{2}=\alpha_2/\alpha_1$ and $\tilde{\alpha}_{NL}=\alpha_{NL}/\alpha_1$. The main advantage of moving to the dimensionless coordinates is that the above Eq.~(\ref{SGPE}) is independent of $\alpha_1$ and we need to vary only two non-linear coefficients $\tilde{\alpha}_{2}$ and $\tilde{\alpha}_{NL}$ in order to check the dependence of polariton propagation on all the three non-linear terms. It is quite easy to see that changing $\alpha_1$ means renormalizing the polariton intensity (in physical unit) as $|\psi_{\sigma_\pm}|^2=E_u |\tilde{\psi}_{\sigma_\pm}|^2/\alpha_1$. Polariton propagation for different cases are shown in Fig.~ \ref{SNonlinear_para_Dep}, which shows that the obtained non-reciprocal effects are quite tolerant to variations in the nonlinear coefficients.

\begin{figure}[H]
\centering
\includegraphics[width=0.5\textwidth]{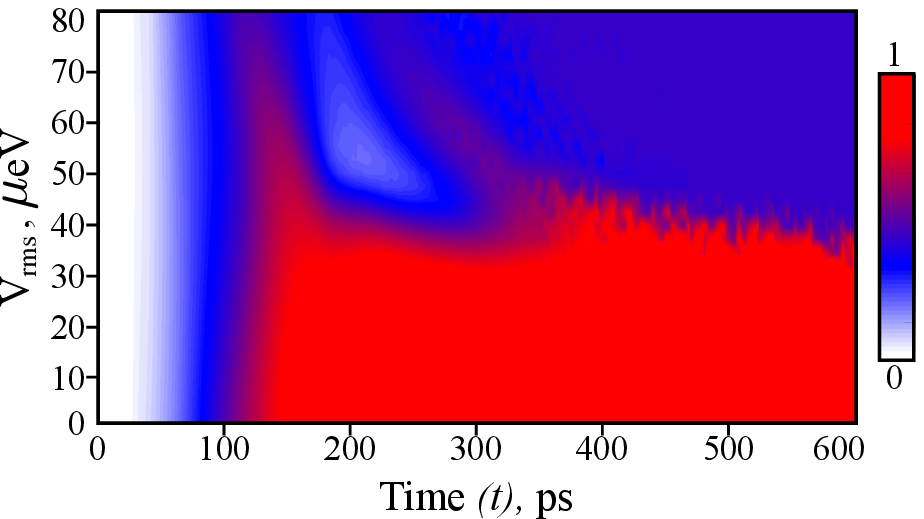}
\caption{$I$ as a function of time and disorder strength. For $V_{rms}<40~\mu$eV, the polaritons reaching the left end do not backscatter resulting in $I \simeq$ 1. However, for larger disorder values $I <$ 1, indicating the breakdown of non-reciprocity.  The parameters of the pulse are kept the same as the one in Fig.~(4) in the main text.}
\label{Fig5}
\end{figure}

{\textit{Effect of spatial disorder.}---}  Disorder is always present in realistic systems. Consequently, to take the disorder into account we add a continuous disorder potential  (characterized by its root mean square value and correlation length) to the system. The robustness of the system is characterized by the quantity, $I={I_L}/{I_T}$, where $I_L$ is the intensity at the left end pillar and $I_T$ is the total intensity of the system. In Fig.~\ref{Fig5}  $I$ is plotted as a function of time $t$ and disorder strength $V_{rms}$, where for each disorder realization an incoherent pulse is launched in the middle of the chain. The white region indicates times for which polaritons did not yet reach the left end. At larger times most of the intensity of the system is located at the left end indicating the non-reciprocal nature. For larger disorder values $I$ decreases, representing comparatively lesser polaritons reaching the left end. However, the typical disorder strength in modern samples ranges between 20-30 $\mu$eV \cite{Heuser_2018,Baboux_2016}, for which the non-reciprocal nature of the system is unhampered. 

\begin{figure}[H]
\centering
\includegraphics[width=0.5\textwidth]{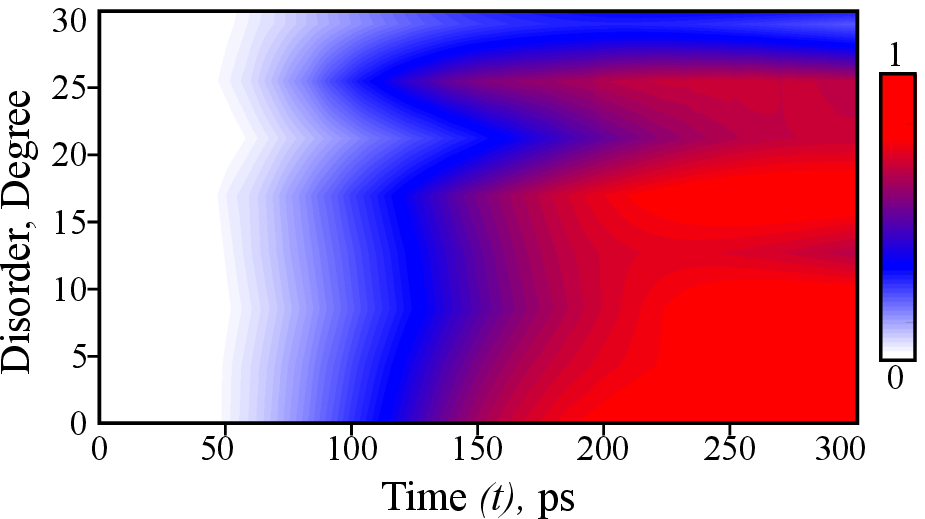}
\caption{$I$ as a function of time and disorder strength. The parameters of the pulse are kept the same as the one in Fig.~(4) in the main text.}
\label{Supp3}
\end{figure}

{\textit{Effect of disorder on the angle between the micropillars}---} In this section we introduce disorder in the angle between the micropillars and to check its robustness we calculate the quantity, $I$, defined above. In Fig.~\ref{Supp3}  $I$ is plotted as a function of time $t$ and disorder strength, where for each disorder realization an incoherent pulse is launched in the middle of the chain. From the figure it is clear that the system can survive disorder up to \ang{25} in the splitting angle. 

\begin{figure}[H]
\centering
\includegraphics[width=0.65\textwidth]{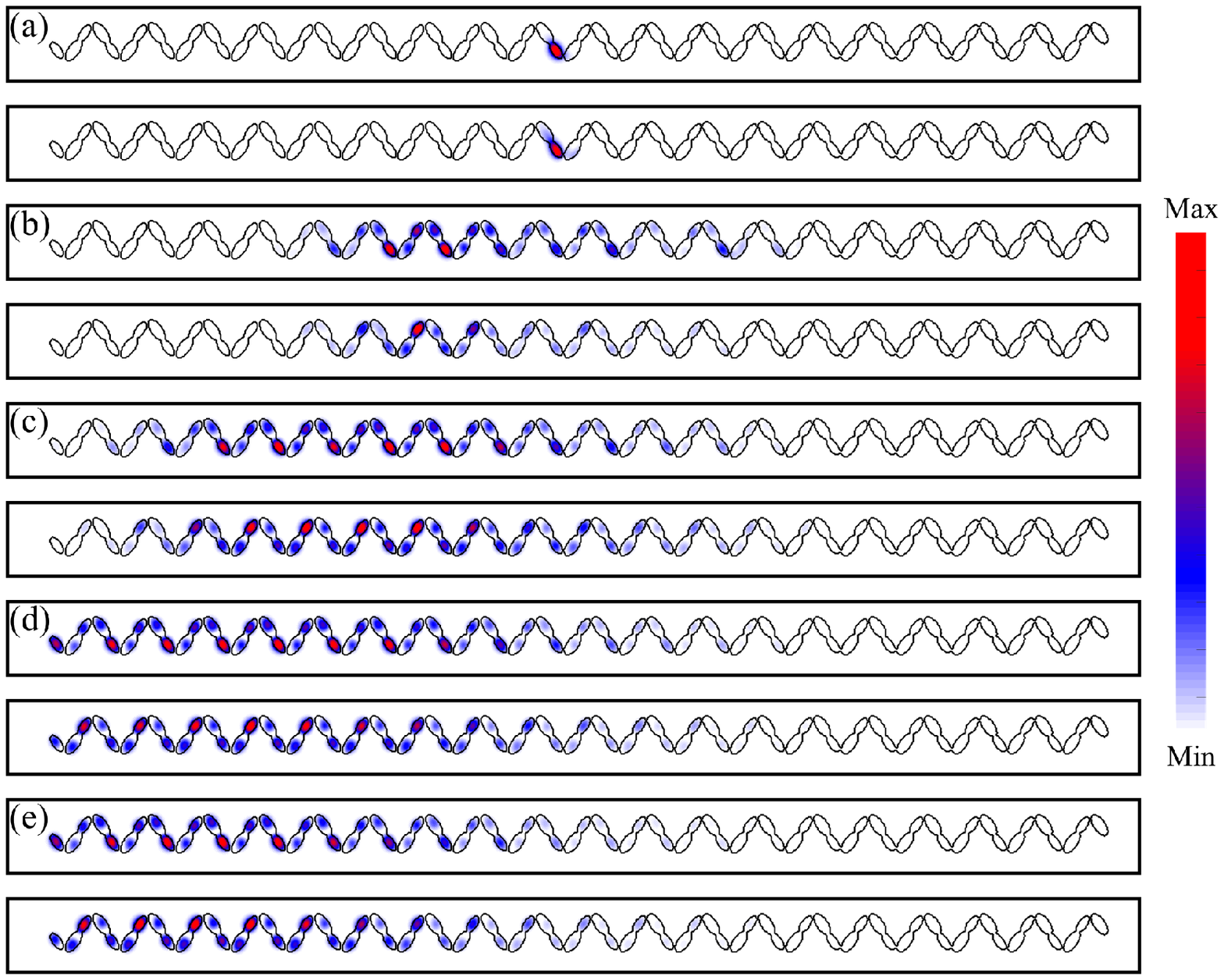}
\caption{Polariton propagation under an incoherent pulse using the coupled reservoir model. (a-f) Intensity of the polaritons ($|\psi_{\sigma_+}|^2$ for top panel and $|\psi_{\sigma_-}|^2$ for bottom panel ) for 10 ps, 50 ps, 100 ps, 150 ps and 200 ps, respectively.  Similar to the main text the slow components of the on-site modes get excited, and they are mostly localized in the micropillars with smaller dimension. The amplitude of the pump $\varepsilon_0=93$ ps$^{-1}\mu$m$^{-2}$  and that of the pulse is  10$\varepsilon_0$ and $\eta=0.2$ ps$^{-1}$.}
\label{SProp_Reservoir}
\end{figure}

{\textit{Pulse propagation using the coupled excitonic reservoir model.}---} In  Eq.~(13) in the main text the effect of the excitonic reservoir is taken into account through the incoherent pumping term $iP_{\sigma_{\pm}}(x,y)$, the exciton induced blueshift term $gP_{\sigma_{\pm}}(x,y)$, and the nonlinear decay term $-i\alpha_{NL}|\psi_{\sigma_{\pm}}|^2$. Such a model, where the density of the excitonic reservoir is approximated as its steady state under the assumption that the reservoir decays much faster,  is widely used, especially in configurations under continuous wave excitation. However, instead of relying on the approximation,  we can also consider the excitonic reservoir explicitly , writing a separate equation for its dynamics  coupled to the driven dissipative Gross-Pitaevskii equation \cite{Wouters_Carusotto_2007}. The dynamics of the polaritons can then be expressed as \cite{Kammann_2012}
\begin{align}
i\hbar\frac{\partial\psi_{\sigma_{\pm}}}{\partial t}=&\left[-\frac{\hbar^2\nabla^2}{2m}+V(x,y)-i\hbar\gamma(x,y)+\alpha_1|\psi_{\sigma_{\pm}}|^2+\alpha_2|\psi_{\sigma_{\mp}}|^2\right]\psi_{\sigma_{\pm}}+V_T\left(x,y,\pm\theta \right)\psi_{\sigma_{\mp}}+\left(g_r+i\hbar \frac{R}{2}\right)n_{\sigma_{\pm}}\psi_{\sigma_{\pm}},\label{SEq101}\\
\frac{\partial n_{\sigma_{\pm}} }{\partial t}=&\left[-\gamma_r-R|\psi_{\sigma_{\pm}}|^2\right]n_{\sigma_{\pm}}+\varepsilon_{\sigma_{\pm}}+\varepsilon^p_{\sigma_{\pm}}\exp\left(-\eta t\right).\label{SEq102}
\end{align}
Here $n_{\sigma_{\pm}}$ represents the density of the incoherent exciton reservoir, $\gamma_r=1.5 \gamma=0.36$ ps$^{-1}$ is decay rate of the reservoir. $R=0.01$ ps$^{-1}\mu$m$^2$ is the condensation rate and $g_r=2\alpha_1$ is the interaction between the polaritons and reservoir excitons. $\varepsilon_{\sigma_{\pm}}$ is the incoherent pump of the form
\begin{align}
\varepsilon_{\sigma_{\pm}}=\varepsilon_0\sum_{(x_0,y_0)}\exp\left[-\left(\frac{\left(x-x_0\right)^2}{w^2}+\frac{\left(y-y_0\right)^2}{w^2}\right)\right],\label{SEq103}
\end{align}
where $\varepsilon_0$ is the amplitude of the pumps and $(x_0,y_0)$ are the position of the pumps having width $w$. The amplitude of the pumps $\varepsilon_0$ is chosen in such a way that it is  just below the condensation threshold and the spatial profile of the pumps are kept the same as those shown in Fig.~3(a-b) in the main text. The last term in Eq.~(\ref{SEq102}) represents an incoherent pulse positioned at the middle of the chain. All the above mentioned parameters related to the characteristics of the excitonic reservoir are taken from Ref.~\cite{MaEgorovSchumacher_2017} and rest of the parameters are kept the same as those used in Fig.~(4) in the main text. Next we solve the coupled  Eqs.~(\ref{SEq101}-\ref{SEq102}) with small random noise as an initial condition for the polaritons and zeros as initial condition for the reservoir. In Fig.~\ref{SProp_Reservoir} the propagation  of the polaritons is shown. Similar to the Fig.~4 in the main text, the polaritons propagate only in one direction with similar group velocity. This proves the robustness of the scheme presented, which does not depend upon the model used to describe the polariton motion.

\textit{Demonstration of the feedback suppression using the coupled excitonic reservoir model.}---
\begin{figure}[H]
\centering
\includegraphics[width=0.75\textwidth]{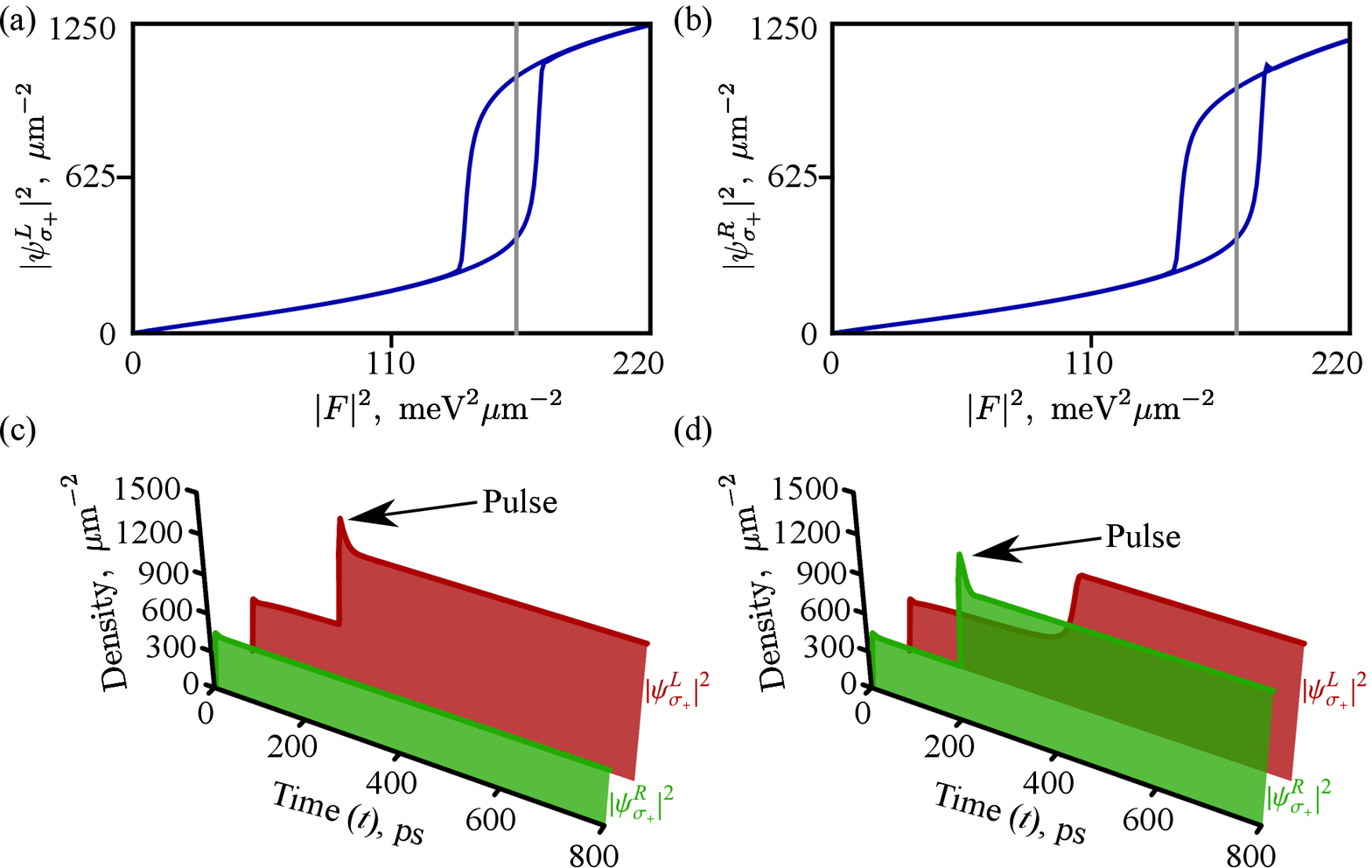}
\caption{(a-b) Bistability curves of the left and right end pillars under coherent pump. (c) Incoherent pulse positioned at the left end does not switch the pillar at the right end. (d) The same pulse positioned at the right end switches the pillar at the left end. The gray line in (a-b) indicates the value of the coherent pump used for demonstration of the feedback free information transfer. The energy of the coherent pump is taken as 1.7 meV and the the amplitude of the pulse is taken as 35$\varepsilon_0$. All other parameters are same as those in Fig.~\ref{SProp_Reservoir}.}
\label{SBis_Reservoir}
\end{figure}
In this section we show feedback free transfer of binary information using the coupled excitonic reservoir model. We add a coherent pump term $F(x,y)$ at the right hand side of Eq.~(\ref{SEq101}), which is positioned at the two ends of the chain. The bistability curves shown in Fig.~\ref{SBis_Reservoir} (a-b) are obtained by slowly varying $F$ with time. Next we fix the pump such that both the end pillars are in the bistable regime. To show the transfer of binary information an incoherent pulse is  launched at the left end. The pulse switches the left end pillar from the lower bistable state to the upper one. Due to the non-reciprocal nature of the system polaritons are not allowed to propagate from left to right and hence no switching is observed for the pillar at the right end (see Fig.~\ref{SBis_Reservoir}(c)). Next, the same incoherent pulse is launched at the right end which switches the right end pillar from the lower bistable state to the upper one. Since, polaritons are allowed to propagate from right to left, after some time the pillar at the left end also switches  to the upper bistable state. In this way the feedback free transfer of bistable information can also be realized by taking into account the dynamics of  the excitonic reservoir.\\

{\textit{Efficiency of the system}---}
In our calculations we chose a typical decay rate of 0.24 ps$^{-1}$, which corresponds to the polariton lifetime of approximately 2 ps. Although polaritons are lost at this rate, the pump replenishes them and consequently, the strength of the signal arriving at the drain is almost the same as that at the source. While transferring information using the bistability, which effectively digitizes the signal, the transmission of information is 100\%. If the bistable states are considered as digital ``0" and ``1", respectively, then switching the state at the input to ``1" will result in switching the state at the output to ``1".

Theoretically, it is challenging to estimate the required external pump power, which depends on the efficiency of light-matter coupling and the efficiency of energy relaxation processes resulting in polariton gain. However, given that our pump power should overcome the losses in the system, the required pump power is similar to that required for polariton lasing. Experiments with a similar polariton linewidth \cite{Bajoni_Wertz_2008} have reported polariton lasing with pumping power in the mW range for a single micropillar of similar size.

The main source of dissipation in the exciton-polariton system is the radiative decay with the rate defined by the lifetime of the cavity photons and the Hopfield coefficient (i.e., proportion of the photon in the polariton). The photon lifetime is limited by the imperfect reflectivity of the microcavity mirrors, and side losses through the edges of the micropillars. Non-radiative loss processes, such as non-radiative exciton recombination occur at the rates several order of magnitude slower (the lifetime of excitons is typically around 1 ns, while the cavity photon lifetime is in the ps range). Phenomenologically, the linear decay rate in the model accounts for the dominant radiative losses.

In experiments with exciton-polaritons, the optical pump is either pulsed or chopped with an AOM to minimize the local heating of the sample, which can manifest itself in redshift of the polariton energy. Our modelling is performed under the assumption that this is done to eliminate the heating effect.  \\

{\textit{Material realization}---}We require elliptical pillars to introduce a linear polarization splitting of exciton-polaritons in individual micropillars, so as to couple modes of different circular polarization. The size of such splitting is the parameter $\Delta_T$. Given the formula for $H_{lr}$  we see that such a splitting is required to have an effective coupling between the modes. While such a splitting could also be introduced through the use of anisotropic materials, we also require the orientation of the splitting to be different in different micropillars (this is the condition that $\theta_l$ differs from $\theta_r$). Elliptical pillars allow us to engineer the polarization splitting with a direction aligned with the pillar orientation.

For material realization, we refer the reader to the experimental study of Ref.~\cite{Klaas_Egorov_2019}. In this work, elliptical pillars were fabricated in microcavities composed of AlGaAs/AlAs distributed Bragg reflector mirrors with embedded GaAs quantum wells. In this system the polarization splitting is controlled by the ellipticity of the pillars, as required for our proposal. In principle, microcavities containing other materials, such as perovskites \cite{Su_Ghosh_2020} and organic materials \cite{Dusel_Betzold_2020} can also be used, as long as the lateral confinement of polaritons in a periodic potential landscape can be engineered along with the orientation-dependent polarization splitting.